\documentclass[a4paper,11pt]{revtex4}
\usepackage{latexsym}
\usepackage{graphicx}
\usepackage[latin1]{inputenc}
\usepackage{amssymb}
\usepackage{graphicx} 

\newcommand{\be}{\begin{equation}}
\newcommand{\ee}{\end{equation}}
\newcommand{\ba}{\begin{eqnarray}}
\newcommand{\ea}{\end{eqnarray}}
\begin{document}
%\begin{frontmatter}



\title{Dynamics  of an inchworm nano--walker}

\author{A. Ciudad\dag\footnote[3]{To whom correspondence should be addressed (aciudad@ecm.ub.es)}, J.M. Sancho\dag\     and A.M. Lacasta\ddag\
}
\address{
\dag\ Departament d'Estructura i Constituents de la Mat\`eria,
Facultat de F\'{\i}sica, Universitat de Barcelona,
Diagonal 647, E-08028 Barcelona, Spain\\
}
\address{
\ddag\
Departament de F\'{\i}sica Aplicada, Universitat Polit\`ecnica de Catalunya,
Av.Dr. Mara\~{n}\'on 44, E-08028 Barcelona, Spain\\
}

\begin{abstract}
% Text of abstract
An inchworm processive mechanism is proposed to explain the motion of dimeric 
molecular motors such as kinesin.We present here preliminary results for this mechanism focusing on observables like mean velocity, coupling 
ratio and efficiency versus ATP concentration and the external load $F$.
\end{abstract}

%\begin{keyword}
%inchworm mechanism \sep molecular motor 

\maketitle
Molecular motor proteins transform the energy of ATP hydrolysis into mechanical 
work performing discrete steps  along a periodic track.  The experimental work on protein motors 
\cite{Visscher,hua,block}  has stimulated a wide 
variety of  modelizations,  most of them based in ratchet--like potentials. The two main candidates for the walking mechanisms of dimeric motors were inchworm or hand--over--hand. 
In the first case it is assumed that the first leading head advances one step which 
is 
followed instantaneously by the trailing head. In the hand--over--hand mechanism the 
second head 
advances two steps overpassing the first head.
In Ref. \cite{hua} some experimental evidence was presented which seem to support 
the 
inchworm mechanism. Nevertheless, more precise experiments show that
myosin--V walk in a hand--over--hand way \cite{block}. These two 
different mechanisms imply different conformational changes in the protein 
structure during ATP hydrolysis. Moreover they imply a different response with 
respect the experimental control parameters [ATP] and $F$. 

Although it is commonly accepted now  that some processive members of 
kinesin, myosin and dynein families seem to walk  in a 
hand-over-hand fashion, it is still worth analyzing  the inchworm mechanism, which 
could hold for other type of motors. For these reasons,  we will present here a 
very simple model walking in a inchworm fashion with parameter values in the   
biological scale. We will also focus on some implications with experimental 
relevance.\\
It was showed in Ref.\cite{Visscher}  that kinesin uses a single ATP  
molecule to perform each step. Such relation is called the
coupling ratio, which for low external loads is about 1.  The temporal distribution 
of these steps is random due to the ATP diffusion until it  reaches the motor. 
After binding the nucleotide, hydrolysis and the consequent  conformational change 
take place displacing the whole motor a certain distance which  is usually equal to 
the periodicity of the track. All this process can occur even in  the presence of 
an opposing external force $F$  and at low ATP concentration, although  both 
regimes 
decrease the mean velocity. \\
The inchworm walking mechanism can be modeled as two linearly-coupled  particles 
interacting with a ratchet potential. The conformational cycle is  introduced as a 
stretching and posterior relaxing of the coupling spring. This way  of modeling was 
introduced in Ref.\cite{tsironis} showing that thermal fluctuations  are not 
strictly necessary in order to achieve the motion. Other works  
\cite{dan,metzler,klumpp,mogilner,linke} are also based on this approach.  The main 
difference between them is  the way they model the mechanical changes under 
the input of chemical energy.  While \cite{tsironis} considers the conformational  
change as an increase of the equilibrium length of the spring, other literature  introduce 
asymmetric frictions or switches on the ratchet potential. Here we  will explore a 
different and simple way which allows to control with precision the  amount of the 
input energy. Furthermore, we apply a kinetic methodology\cite{inhibition}  based 
on enzymatic inhibition to 
get an analytical expression for the velocity as a  function of the ATP 
concentration and the external force $F$. Finally we  analyze the coupling ratio 
and the efficiency at different values of $F$. 

\begin{figure}
		\begin{center}
	\includegraphics[width=18 cm,bb=0 0 426 149]{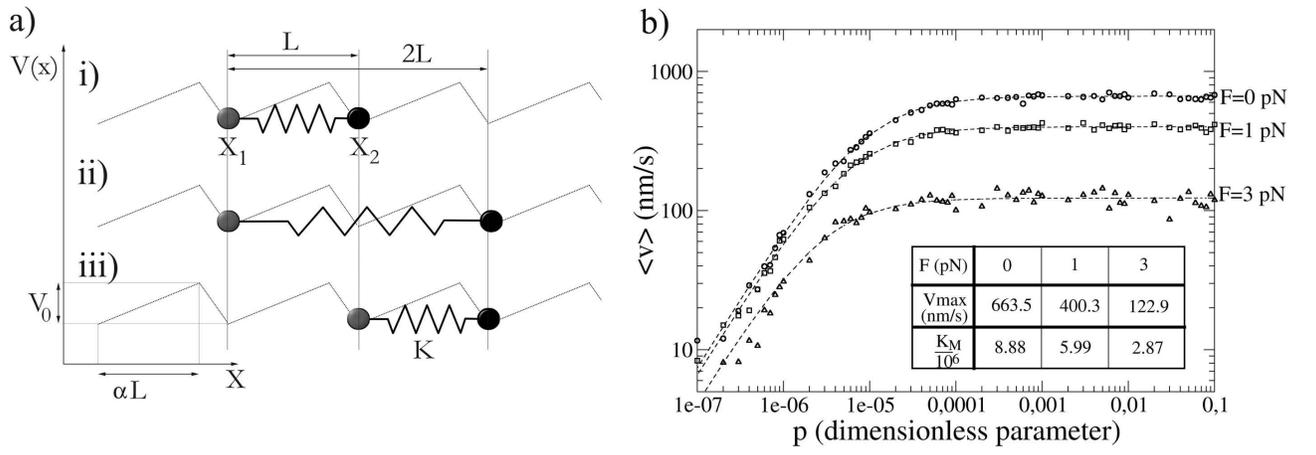}
% one.ps: 72dpi, width=15.03cm, height=5.26cm, bb=0 0 426 149
	\end{center}

	\centering

\caption{ {\bf a)} Scheme of the ratchet potential the positions of the leading head (black particle) and the trailing head (gray particle) at the three stages of the motor. {\bf i)} Rest configuration. {\bf ii)} The motor at the end of the  stretching. {\bf iii)} Final stage.  {\bf b)}
Simulated mean velocities $\langle v\rangle$ of the center of mass versus $p$ for $F=0,1,3$ pN drawn with circles, squares and triangles, respectively.  Dashed lines are 
Michaelis-Menten fits. The resulting kinetic  parameters are shown in the table.
 }
\label{1}
\end{figure}

We will consider the motor as two particles coupled by a spring. The set of 
equations in the overdamped limit are,
\begin{eqnarray}
\lambda\dot{x}_1=-V'(x_1)-k(x_1-x_2-L)-f_s(t)-\frac{F}{2}+\xi_1(t)
\nonumber\\
\lambda\dot{x}_2=-V'(x_2)+k(x_1-x_2-L)+f_s(t)-\frac{F}{2}+\xi_2(t),
\end{eqnarray}
where $x_1$,$x_2$ are the position of the  trailing and the
leading head, respectively. $k$  is the  stiffness of the harmonic spring with  
equilibrium length $L$, which is also the periodicity of the  ratchet 
potential $V(x)$. Such potential has an asymmetric factor $\alpha$ and a  barrier 
height 
$V_0$ (See Fig.\ref{1}). $f_s(t)$ is the random chemical force and $F$ is the 
external load.
The thermal force is emulated through a zero mean   
Gaussian-white 
noise with a correlation, 
$\langle\xi_i(t)\xi_i(t')\rangle=2\lambda k_BT\delta(t-t')$.   $\lambda$ is the 
friction and 
 $k_BT$ is the thermal energy. 
Then, at thermal equilibrium the two  particles will 
lay, most of the time, on two consecutive potential minima. We assume that  at 
random intervals of time, an energetic 
nucleotide like ATP will bind the motor and a stretching force $f_s$  will act on 
the system until the total length of the motor will be doubled, i.e.  $x_2-x_1=2L$. 
If $E$ is the hydrolysis energy of the molecule and $L$ is the  displacement that 
it is performed, then we take the chemical force $f_s=E/L$.\\
The values of the parameters have been chosen in a nano scale to mimic some 
molecular motors such the kinesin.  The periodicity of 
the potential $L$ is taken to be the periodicity of microtubules, $8 nm$.
The asymmetric factor $\alpha=0.8$ and $V_0=50pNnm$
 optimizes the efficiency of our model.  
$E=100pNnm$ corresponds to an accepted value for  the energy of  hydrolysis of 
an ATP and  thermal energy is $k_BT=4.1pNnm$. The stiffness of the motor is  chosen 
$k=1pN/nm$ and the drag force $\lambda=2\cdot 10^{-4}pNs/nm$. 
When the motor is free from ATP, $p\in(0,1)$ is the uniform probability  per time 
step $\Delta t$ of binding one molecule. When it occurs, more ATP binding is  
forbidden and 
stretching takes place until the elongation is $2L$. Then, the stretching  force 
disappears and the spring relaxes.  When $x_2-x_1$ is again $L$, one cycle is 
completed and ATP binding is allowed.  In the absence of external load, this 
mechano-chemical cycle induces a $L$  displacement of the motor towards one end of 
the potential. Fig. \ref{1}a shows the  scheme of the process. 
On the other hand, our approach controls how much energy $E_T$ is 
applied to the system by simply multiplying $E$ by the number $n$ of ATP consumed: $E_T=nE$. The mean velocity of the motor, when the ATP  concentration is saturant and $F=0$, 
is maximum and dependent only on the intrinsic properties  of the motor and by $E$. Let 
$t_{on}$ be the time spent to perform a single  step, i.e. the stretching plus the 
relaxing time. Thus, $V_{max}=L/t_{on}$.  Using the given values of the 
parameters, simulations show that $t_{on}\sim0.012$s,  which gives 
$V_{max}\sim667$nm/s. However, the global speed $\langle  v\rangle$ will be slowed 
down when the ATP concentration decreases.  Typically, the 
$[ATP]$-dependence on $\langle v\rangle$ is given by the Michaelis-Menten relation 
\cite{inhibition}. In our model, we have previously defined $p$ 
as the uniform probability to get an ATP  per time step $\Delta t$ and with $p=0$ 
while the motor stretches and relaxes. It  can be accepted that, as the reaction 
frequency is proportional to $[ATP]$,  and then $[ATP]$ is proportional to $p$. 
Then, we have 
\be
\langle v \rangle =V_{max}\frac{p}{K_M+p},
\label{eMM}
\ee
where $K_M$ is the Michaelis constant for the probability.
From now on, we will deal with $p$ and not with $[ATP]$. Fig.\ref{1}b shows how the michaelian behavior fits well the simulated values of the mean velocity. However, for finite values of $F$, both kinetic parameters $V_{max}$ and $K_M$ change. In Ref.\cite{inhibition} it is shown that the effect of the external load in kinesin can be interpreted as an inhibition process. This introduces a $F$-dependence on the two kinetic parameters 
\be
V_{max}(F)=\frac{V_{max}(F=0)}{1+\frac{1}{K_{iu}(F_S/F-1)}} \qquad K_M(F)=K_M(F=0)\frac{1+\frac{1}{K_{ic}(F_S/F-1)}}{1+\frac{1}{K_{iu}(F_S/F-1)}}.
\label{e3}
\ee
and allows to express the velocity of the motor as a function of the two control variables $p$ and $F$.\\
\begin{figure}
		\begin{center}
	\includegraphics[width=18 cm,bb=0 0 426 156]{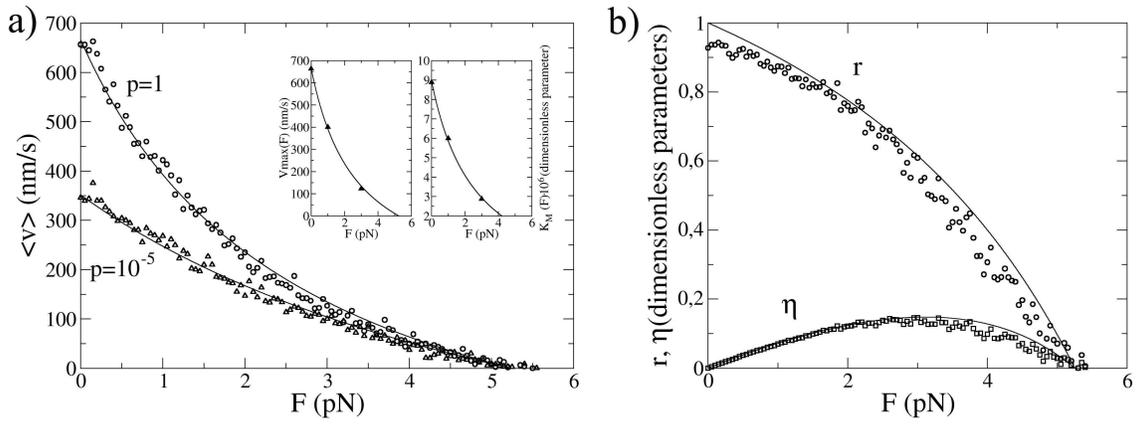}
% two.ps: 72dpi, width=15.03cm, height=5.50cm, bb=0 0 426 156
	\end{center}

\caption{{\bf a)} Simulated mean velocities versus  $F$ for $p=1$ (circles) and 
$p=10^{-5}$ (triangles). The insets show $V_{max}(F)$ and $K_M(F)$ versus $F$. Triangles  are the values from the table in \ref{1}b). Solid lines are the fits of (\ref{e3})  in order to get $K_{iu}$ and $K_{ic}$. With the two inhibition constants, (\ref{eMM}) can be plotted obtaining the solid lines of the main figure.
{\bf b)} Coupling ratio $r$ and efficiency $\eta$. Circles are simulated data and solid lines are predictions of (\ref{e4}).
}
\label{2}
\end{figure}
$F_S$ is the stall force, i.e. the maximum load that  the motor is able to carry. 
In our simulations, $F_S\sim5.25$pN. $K_{iu}$ and  $K_{ic}$ are the uncompetitive 
and competitive inhibition constants, respectively,  and are a quantitative measure 
of how $F$ affects the motor when it is free from  nucleotide ($K_{ic}$) or when it 
has an ATP ($K_{iu}$). From table values of Fig.\ref{1}b  we fit the values of the 
inhibition constants obtaining $K_{iu}\sim0.338\cdot10^{-6}$ and $K_{ic}\sim2.131\cdot10^{-6}$. As they are dissociation constants, the effect of the load on the ATP-bound state is greater than in the ATP-free configuration. This means that the force acts as an uncompetitive mixed inhibitor, while in Ref.\cite{inhibition} it is shown that kinesin is also mixed but competitive. This difference is responsible of the curvature on $\langle v\rangle-F$ curves at high ATP concentration. Figure \ref{2}a shows these curves with the simulation data and the predictions of the analytical expression with an excellent agreement.\\
Finally, it is interesting to define the coupling ratio and the efficiency and to see how are they modified by $F$. The coupling ratio $r$ can be expressed as the quotient between the total number of performed steps and the total number of consumed ATP's. On the other hand, the efficiency $\eta$ can be defined as the ratio between the work performed against $F$, $W$, and the total input of energy, $nE$. Thus,
\be
r\equiv\frac{x_{CM}}{nL} \qquad \qquad \eta\equiv\frac{W}{nE},
\ee
where $x_{CM}=\frac{1}{2}(x_1+x_2)$ and supposing that $x_{CM}(t=0)=0$.
$W=Fx_{CM}$, so we can write $\eta=rFL/E.$
This means that the global efficiency is  simply the efficiency in a single step 
multiplied by the coupling ratio. We can go further if we consider (\ref{e3}) and 
the fact that $V_{max}$ is proportional to $r$, and then,
\be
r=\frac{1}{1+\frac{1}{K_{iu}(F_S/F-1)}} \qquad\qquad \eta=\frac{L}{E}\frac{F}{(1+\frac{F}{K_{iu}(F_s-F)})}.
\label{e4}
\ee
Fig.\ref{2}b shows the simulated data for $r(F)$ and $\eta(F)$ as well as the theoretical predictions. It is interesting to remark that the maximum efficiency is slightly below $0.15$.

We have presented an inchworm mechanism which is able to perform directed transport and analyzed how it behaves under two variables, the ATP concentration through the probability $p$ and the external load $F$. The motor can be described as a uncompetitive mixed inhibitor obtaining analytical expressions for the mean velocity as a function of the two control variables that fits accurately the simulated data. Finally, we have discussed the coupling ratio, showing that the motor loses the tight coupling as $F$ increases. The efficiency is related with the coupling ratio and an analytical expression is given with a good agreement with the simulations.\\
This work was supported by the Ministerio de Educaci\'on y Ciencia 
(Spain) under the project No. $BFM2003-07850$ and the grant No. 
$BES-2004-3208$(A.C.). 
{}
\end{document}